# Understanding the Changing Landscape of Automotive Software Vulnerabilities: Insights from a Seven-Year Analysis

1st Srijita Basu
*Computer Science & Engineering*
*Chalmers University of Technology*
*and University of Gothenburg*
417 56 Göteborg, Sweden
srijita.basu@gu.se

2nd Miroslaw Staron
*Computer Science & Engineering*
*Chalmers University of Technology*
*and University of Gothenburg*
417 56 Göteborg, Sweden
miroslaw.staron@gu.se

*Abstract*—The automotive industry has experienced a drastic transformation in the past few years when vehicles got connected to the internet. Nowadays, connected vehicles require complex architecture and interdependent functionalities, facilitating modern lifestyles and their needs. As a result, automotive software has shifted from "just embedded system/SoC (System on Chip)" to a more hybrid platform, which includes software for web/mobile applications, cloud, simulation, infotainment, etc. Automatically, the security concerns for automotive software have also developed accordingly. This paper presents a study on automotive vulnerabilities from 2018 to September 2024, i.e., the last seven years, intending to understand and report the noticeable changes in their pattern/trend. 1,663 automotive software vulnerabilities were found to have been reported in the studied time frame. The study reveals the Common Weakness Enumeration (CWE) associated with these vulnerabilities develop over time and how different parts of the automotive ecosystem are exposed to these CWEs. Our study provides the platform to understand the automotive software weaknesses and loopholes and paves the way for identifying the phases in the software development lifecycle where the vulnerability was introduced. Our findings are a step forward to support vulnerability management in automotive software across its entire life cycle.

## I. INTRODUCTION

The ability to exploit the different vulnerabilities in automotive software is not only a threat to the reliability of the software and its functionalities but also to the vast amount of data (associated with the Original Equipment Manufacturer (OEM), supplier, occupant, etc.) processed by this software. The different instances of ransomware attacks, service outages, and data leakages reported in the last few years for some of the most renowned automotive manufacturers (like Tesla, Honda, Toyota, etc.) [1] [2] [3] prove this fact. Identifying vulnerabilities in the development phase alone is not enough for automotive software. The entire software engineering team should have a holistic view and follow proper security tactics to ensure vulnerability identification and mitigation across all phases of the Software Development Lifecycle (SDLC). Researchers need proper automotive data and statistics to conceptualize an adequate design strategy for automotive-centric security techniques or tools.

The Common Vulnerabilities and Exposures (CVE) repository [4] by the National Vulnerability Database (NVD) is a useful source for both industry and academia to get an insight of the data and overall trend of security vulnerabilities across multiple domains. However, for any data-driven vulnerability research that focuses on a specific domain, like the automotive domain, getting a filtered view of this huge CVE repository is necessary. The view and proper trend analysis of the automotive vulnerability records over a moderate time frame would reveal the actual picture.

In this paper, we conduct an empirical study on the evolution of automotive vulnerabilities over 2018 to 2024. It should be mentioned in this context that though we had NVD data available for previous years, we chose to start our study from the year 2018 as connected cars became fully functional and gained popularity during this time. As a part of this work we identified four research questions (RQs). In the process of answering them, we present some interesting findings on automotive vulnerabilities over the past seven years. These could further facilitate security researchers' exploration of automotive software security trends and the development of mitigation strategies accordingly. The RQs are as follows:
- **RQ1**: How did the number of reported automotive

This research has been partially funded by Strategic Research Area Transport, University of Gothenburg

vulnerabilities change over the last 7 years?
- **RQ2**: How did the distribution of CVSS for automotive vulnerabilities change in the last 7 years?'
- **RQ3**: How did the trend of attack vectors for automotive vulnerabilities change in the last 7 years?
- **RQ4**: How did the trend of CWEs change for automotive vulnerabilities in the last 7 years?

The paper makes the following contributions in the context of automotive vulnerabilities for the period of 2018 – 2024 September:
- Identify the change in the distribution of Common Vulnerability Scoring System (CVSS) [5] across these seven years
- Map the vulnerabilities with attack vectors, i.e., identify the software component that introduced the vulnerability into the system. Also, report any change in trend associated with this mapping
- Present the existing mapping of the vulnerabilities with their respective Common Weakness Enumeration (CWEs) [6] to reflect the changes in trend/pattern
- Share some reflections on how to use this data for mapping the vulnerabilities to the SDLC phases in which they were introduced

The rest of the paper is organized as follows. Section II presents the literature background. Section III describes the CV ecosystem focusing on automotive software components. Section IV details the methodology for filtering automotive vulnerability records and mapping procedures. Section V provides insights into the study's results, their analysis, and some discussion. Section VI describes the validity of our work. Finally, Section VII concludes the paper.

## II. Related Work

We surveyed research works that deal with data-driven vulnerability studies, more specifically automotive vulnerabilities. A few of them have been presented here.

The authors in [7], propose an automated tool for collection of CVEs and their fixes from open source software repositories. The authors also presented some analysis (average CVSS, exploitability, impact, etc.) on the vulnerability data that they had collected for the span of 2002-2021. The work is beneficial in the sense that the study could be easily extended after 2021 but the limitation is, it considers only those CVEs that are associated to open source software.

In Xiong et al. [8], the authors investigate automotive vulnerabilities for 60 OEMs from the NVD and analyze the corresponding CVEs in context of CWE and CVSS. The period of the study was not quite clear and the results provided an overview of all the CWEs and CVSS base metrics that were encountered. The work lacked a reflection of these results on the automotive industry. The authors also suggest some generic mitigation techniques based on some of the most frequently occurring CWEs.

Gülsever [9] presents automotive vulnerabilities for the period, 2010-2019. The data was collected from the NVD and Upstream repositories. The authors identified some attack vectors and provided a count for each (i.e. number of instances when a vulnerability was caused for this attack vector). Some of the most popular CWEs identified in this context were also presented. The percentage of physical and remote attacks possible, based on the CVSS base metrics was also discussed here. The method of enumerating the attack vectors was not clear from the paper. Moreover, the change in trend from 2009-2019 was not very clear from the details that the authors have provided.

Software bugs in automotive software of Connected Vehicles (CV) was reported and analyzed by Garcia et al. in [10]. A discussion on commits and bugs associated with Baidu Apollo and Autoware, two of the most popular open-source CV communities was depicted in this paper. The analysis discovers the root cause of these bugs and identifies the most commonly affected software components. Finally, the authors provide recommendations regarding the detection, localization, and repairs of the automotive software bugs.

The existing studies and analyses on vulnerabilities and security concerns of automotive software are quite interesting. Though, we get a broader overview of the CV security landscape from them, a more focused approach leading to the realization of the major changes that have affected the security of automotive software in the past years is missing. In this paper, we try to bridge this gap by proposing a data-driven vulnerability study for automotive software, spanning a period of seven years. This would not only help us understand the changing landscape of automotive software vulnerabilities but also provide some facts useful for the SE team in secure software development process.

## III. Software Ecosystem for Connected Vehicles

The Connected Vehicle (CV) ecosystem is vast and varied. Before getting into the vulnerability study of automotive software, it is important to understand the notion of the same in the context of the CV ecosystem. This idea is necessary as it would help us filter the vulnerabilities conceptually in some cases, i.e., logically decide whether a particular vulnerability can be associated with automotive software. Fig. 1 presents a compact view of the CV ecosystem and shows the placement of different automotive software.

According to Fig. 1, the CV ecosystem has been broadly divided into the following four parts:
- *In-Vehicle Ecosystem*: Previously, embedded vehicle software was the only part of this ecosystem. However, separate software responsible for each CV functionality can be found here. A few of these software are Telematics, Infotainment, Remote Keyless System (RKS), Sensor and GPS, and Electronic Control

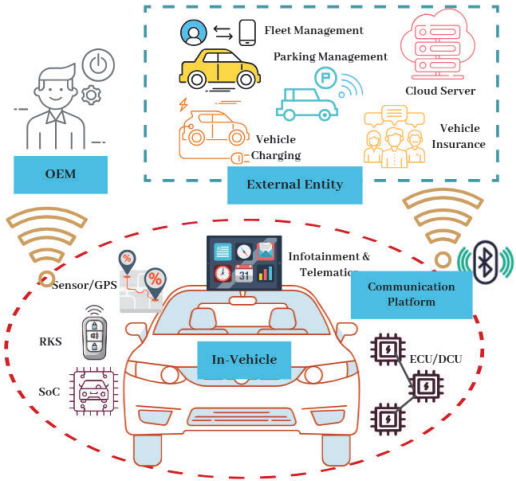

Fig. 1. Automotive Software: Connected Vehicle Ecosystem

Unit (ECU). The System-On-Chip (SoC)/Embedded system still exists, and some of the major vehicle functionalities are programmed within it.
- *External Entity Ecosystem*: Any external system other than the OEM that communicates with the *In-Vehicle Ecosystem* falls under this part. This can contain a wide range of software like Vehicle Charging System, Traffic Management, Parking System, Fleet Management, Car Rental, Car Dealership, Vehicle Insurance, etc.
- *OEM/Supplier Ecosystem*: This part consists of different orchestration tools, build tools, container software, simulation software, etc., used for developing, testing, and maintaining the CVs.
- *Communication Platform*: This consists of the different communication mediums and protocols like Bluetooth, Wi-fi, Dedicated short-range communications (DSRC) [11], MQTT (Message Queuing Telemetry Transport), used for V2X, i.e., Vehicle to Everything communication as well as the OTA (Over-The-Air) updates between the In-Vehicle and OEM ecosystem.

The complex ecosystem requires knowledge of all the software components involved, which is essential for proper security engineering in the automotive domain. With this insight, we filter automotive vulnerabilities and map them to appropriate attack vectors in the next section.

## IV. Methodology: Automotive Vulnerability Filtration, Organization and Mapping

The main methodology of this study can be divided into two sections. First, the vulnerabilities were collected and filtered, and then they were mapped with the corresponding attack vectors. Any manual analysis or decisions involved a group of two researchers, followed by a peer review conducted by another group of two senior researchers. Fig 2 provides an overview of the process involved.

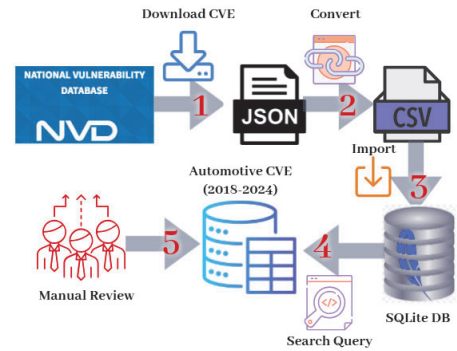

Fig. 2. Method: Automotive CVE filtration

### A. Vulnerability Collection and Filtration

The following steps were involved in the process:

1) *Download CVE List*: Year-wise CVE list (2018 to 2024) is downloaded from NVD in JSON format. The URL[1] is used for the purpose. We can replace `<year>` with the value of the exact year (e.g., 2022) for which we want to download the entire CVE list and associated details
2) *JSON to CSV conversion*: The JSON files are converted into .csv files using *jq* command (JSON processing tool).
3) *Importing the CSV files into Database*: The CSV files are then imported as tables into an SQLite database (A database should be created in SQLite before this step)
4) *Filtering Automotive CVEs - Phase 1*: Once we had the year-wise CVEs as SQL tables, we ran a search query to filter out the automotive vulnerabilities from them. This step required some additional testing. An initial list of 172 automotive keywords (inspired from [18] and considered some recent automotive OEMs and associated terms) was taken to design the search query. This resulted in many false (non-automotive) entries, and many keywords did not contribute to any automotive CVE. Therefore, the search query was fine-tuned in two ways. Firstly, the query was rewritten to reduce the number of false rows as much as possible. For example, the query would return the CVEs whose description contains the word "CAR" but not "CARD". Secondly, the query size was reduced by eliminating the search attributes that did not contribute towards automotive CVEs. For example, the word "ALFA ROMEO" (car manufacturing company) did not return any automotive vulnerability. So, it was removed from the query. A part of the final query is shown below.
5) *Filtering Automotive CVEs - Phase 2*: After getting the filtered automotive CVEs from the previous step,

---

[1]https://nvd.nist.gov/feeds/json/cve/1.1/nvdcve-1.1-⟨year⟩.json.zip

**Automotive CVE Search Query**
SELECT * FROM CVE_2018_21 WHERE field2 LIKE '%AUDI%' AND field2 NOT LIKE '%AUDIO%' AND field2 NOT LIKE '%AUDIT%' AND field2 NOT LIKE '%AUDIE%' AND field2 NOT LIKE '%AUDIB%'AND field2 NOT LIKE'%AUDIM%'OR field2 LIKE '%AUTOLIV%' OR field2 LIKE '%AUTOMOBILE%' OR field2 LIKE '%AUTOMOTIVE%'OR field2 LIKE '%AUTOWARE%' our researchers manually studied the list. A few vulnerabilities were found that contained words like "VEHICLE" but are not automotive vulnerabilities (e.g., CVE-2021-33852- A cross-site scripting (XSS) attack can cause arbitrary code (JavaScript) to run in a user's browser and can use an application as the **vehicle** for the attack.). Such entries were manually removed.

At the end of these steps, we have a list of **1663** automotive vulnerabilities between 2018 and 2024. The CVE list can be accessed from this link [2]. Each data row contains, CVE ID, description, CVSS (denoting the severity of the vulnerability), CWE ID and CWE description. Later we add the attack vector mapping that is described in the next sub-section.

*B. Attack Vector Mapping*

Automotive attack vectors are identified from the CV ecosystem discussed in Section III. Our researchers developed a list of eight attack vectors - SoC/Embedded System, Infotainment, Telematics, ECU, Sensor (part of the in-vehicle ecosystem), External Applications, OEM Software, and Communication Platform. Next, each CVEs was mapped to one of the attack vectors.

The mapping technique was manual in this case. The CVE descriptions were studied, and the appropriate attack vector was selected for the corresponding vulnerability. Table I presents the manual method used for mapping along with examples. Since quite transparent method was used for this mapping and we had used a three layered approach for the same (as depicted in Table I), no disagreements practically occurred in this step.

V. RESULTS, ANALYSIS, AND DISCUSSION

As we discussed the filtration and mapping methodology in the previous section, we are ready to get some insight into this vulnerability data. We explore the trend changes in automotive vulnerability over a period of seven years in the form of answers to our research questions.

**RQ1**: How did the number of reported automotive vulnerabilities change over the last 7 years?

The change in the number of reported automotive vulnerabilities has been shown in Fig. 3. The numbers show that, after a considerable rise in 2019, the number of vulnerabilities has gradually decreased. When further investigated, it was seen that many Qualcomm SoCs were reported with vulnerabilities this year, which resulted in a spike.

[2]https://github.com/SriAbir/AutomotiveCVE_2018_24

TABLE I
METHOD: ATTACK VECTOR MAPPING

| Mapping Method | CVE Description | Attack Vector |
|---|---|---|
| **Direct**- Search for the presence of the attack vector term (e.g., Infotainment) in the CVE description | *CVE-2023-28885*: The MyLink infotainment system (build 2021.3.26) in General Motors Chevrolet Equinox 2021 vehicles allows attackers to cause a denial of service (temporary failure of Media Player functionality) via a crafted MP3 file. [12] | Infotainment |
| **Indirect**- If we get a more specific term in the CVE description (e.g., TeslaMate), try to find the type of this software (e.g., Telematics) | *CVE-2023-29857*: An issue in Teslamate v1.27.1 allows attackers to obtain sensitive information via directly accessing the Teslamate link. [13] | Telematics |
| **Indirect**- Sometimes, the CVE description provides very little information about the associated attack vectors. In such cases, the hyperlink or the Common Platform Enumeration (CPE) associated with the CVE is studied | *CVE-2023-33031*: Memory corruption in Automotive Audio while copying data from ADSP shared buffer to the VOC packet data buffer. [14] | SoC/Embedded System |

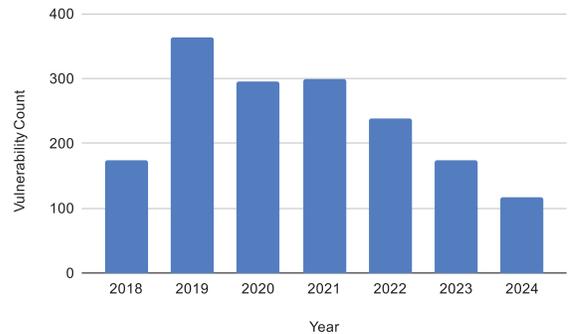

Fig. 3. Count of Automotive Vulnerabilities (2018-2024)

**RQ2**: How did the distribution of CVSS for automotive vulnerabilities change in the last 7 years?

Fig. 4 reflects the CVSS distribution. Since the amount of data was substantial, we used a violin plot to understand the density and distribution of this data. The mean CVSS value ranges from 7.19 in 2024 (lowest) to 8.08 in 2019 (highest), which highlights the fact that most of the vulnerabilities have **HIGH** severity score (A CVSS score within 7.0-8.9 implies HIGH severity).

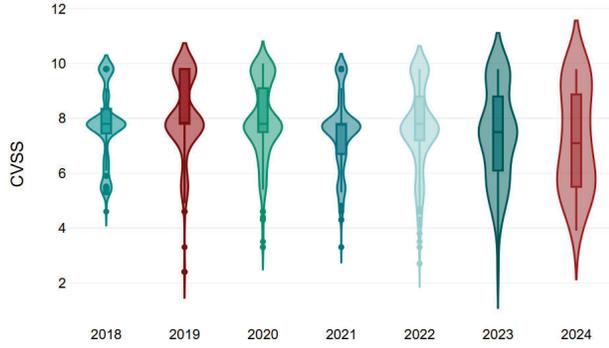

Fig. 4. Automotive CVSS Distribution (2018-2024)

If we look at the CVSS data density distribution per year, it is evident that 2018 has the maximum number of CVSS concentrated within the range of 7.5 to 8.5. So, there is less variation in values. If we look at the curves from 2021 to 2024, this range's size has gradually increased. It is maximum in the case of 2024, where most CVSS values are distributed between 5.5 and 8.9. This indirectly signifies that the vulnerabilities reported in the year 2018 had fewer variations (they were of similar nature, i.e., sharing the same attack vector and/or CPE and/or CWE). However, 2024 will have a large variation in the nature of automotive CVEs. Moreover, it is also evident that this variation started in 2021.

**RQ3**: How did the trend of attack vectors for automotive vulnerabilities change in the last 7 years?

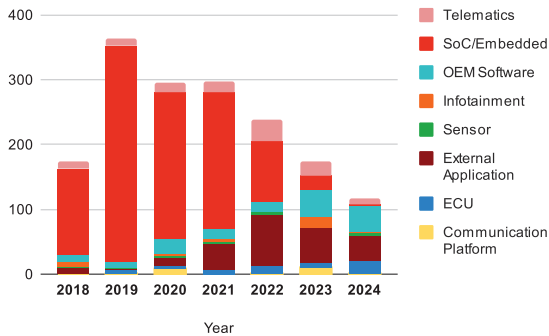

Fig. 5. Automotive Attack Vector Trend (2018-2024)

The attack vector distribution for automotive vulnerabilities is presented in Fig. 5. The most significant change observed is the shift of attack vectors from SoC/Embedded system software to external applications. Another impression was that the reporting of OEM software related vulnerabilities is gradually increasing. For concrete and better visualization, we divided the studied time frame into two parts: 1) 2018-2021 and 2) 2022-204. Fig. 6 shows each automotive ecosystem's vulnerabilities percentage in the divided timelines. We observed a decrease in the in-vehicle vulnerability percentage and an increase in the external and OEM vulnerability percentages over time. There was also a subtle increase in the vulnerabilities associated with the communication platform.

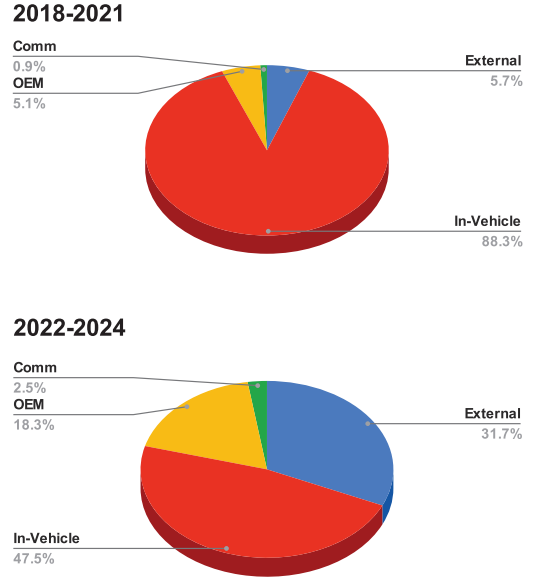

Fig. 6. Comparison of Automotive Attack Vector Trend

The exact count of these attack vectors i.e. number of vulnerabilities caused by a specific attack vector per year is presented in Table II. Looking in to the details, it is observed that *SoC/Embedded* alone shows a decreasing trend over time. The attack vectors, *External Application, ECU, Infotainment, OEM Software and Telematics* shows an increasing trend (the count for 2024 might be less as the entire year could not be considered).

TABLE II
Attack Vector Count for Automotive Vulnerabilities (2018-2024)

| Attack Vector | 2018 | 2019 | 2020 | 2021 | 2022 | 2023 | 2024 |
|---|---|---|---|---|---|---|---|
| Communication Platform | 0 | 2 | 8 | 0 | 1 | 10 | 2 |
| ECU | 2 | 4 | 6 | 6 | 13 | 8 | 20 |
| External Application | 9 | 3 | 11 | 41 | 77 | 53 | 38 |
| Infotainment | 7 | 1 | 3 | 5 | 0 | 17 | 3 |
| OEM | 10 | 8 | 23 | 17 | 15 | 41 | 41 |
| Sensor | 2 | 1 | 4 | 3 | 5 | 1 | 3 |
| SoC/Embedded | 134 | 334 | 226 | 209 | 95 | 22 | 1 |
| Telematics | 11 | 10 | 14 | 16 | 33 | 22 | 9 |

**RQ4**: How did the trend of CWEs change for automotive vulnerabilities in the last 7 years?

We identified 12 CWEs predominantly visible in automotive vulnerabilities across the studied timeline (we used the existing mapping present in NVD). The identified CWEs and their year-wise count of occurrences are presented in Table III. The highest occurrences were found for CWE-120 (51), CWE-125 (47), and CWE-787 (47) in 2019. These issues are common to SoC/embedded vehicle software.

TABLE III
CWE Count for Automotive Vulnerabilities (2018-2024)

| CWE | 2018 | 2019 | 2020 | 2021 | 2022 | 2023 | 2024 |
|---|---|---|---|---|---|---|---|
| CWE-119 (Improper Restriction of Operations within the Bounds of a Memory Buffer) | 30 | 25 | 11 | 4 | 8 | 0 | 2 |
| CWE-129 (Improper Validation of Array Index) | 13 | 18 | 16 | 6 | 6 | 0 | 0 |
| CWE-416 (Use After Free) | 11 | 30 | 11 | 15 | 7 | 2 | 1 |
| CWE-120 (Buffer Copy without Checking Size of Input ('Classic Buffer Overflow')) | 2 | 51 | 26 | 30 | 11 | 3 | 2 |
| CWE-125 (Out-of-bounds Read) | 10 | 47 | 37 | 26 | 19 | 5 | 1 |
| CWE-190 (Integer Overflow or Wraparound) | 2 | 18 | 14 | 14 | 9 | 1 | 0 |
| CWE-20 (Improper Input Validation) | 12 | 14 | 22 | 16 | 3 | 2 | 9 |
| CWE-476 (NULL Pointer Dereference) | 0 | 19 | 5 | 16 | 3 | 1 | 4 |
| CWE-787 (Out-of-bounds Write) | 9 | 47 | 30 | 21 | 12 | 18 | 0 |
| CWE-617 (Reachable Assertion) | 0 | 2 | 8 | 20 | 3 | 0 | 0 |
| CWE-79 (Improper Neutralization of Input During Web Page Generation ('Cross-site Scripting')) | 9 | 2 | 8 | 16 | 17 | 28 | 10 |
| CWE-89 (Improper Neutralization of Special Elements used in an SQL Command ('SQL Injection')) | 2 | 1 | 3 | 4 | 40 | 18 | 19 |

Fig. 7 shows the box plot for the CWE counts. It is evident from the plot that CWE-125 has the highest mean value (20.7), followed by CWE-787 (19.6). CWE-617 (4.7) holds the lowest position. This implies that CWE-125 and CWE-787 are the most frequently occurring CWEs in the last 7 years, and although CWE-617 is fairly visible in the list, its influence is comparatively low. Again, it is also evident from the plot that, CWE-120 and CWE-125 are spread more evenly across the studied time span than the other CWEs.

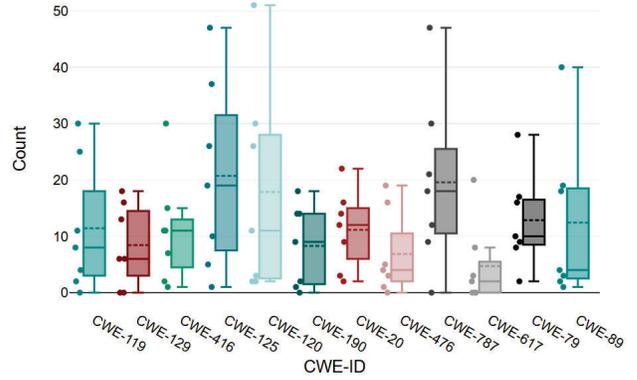

Fig. 7. Automotive CWE Counts (2018-2024)

Next, we tried to understand the trend of increase/decrease in the count of these CWEs. We identified 2 groups of CWEs in this process: i) Increasing Trend and ii) Decreasing Trend. Some of these have been shown in Fig. 8. The upper part of the diagram shows CWEs whose counts gradually decrease with the year. On the other hand, the lower part shows the CWEs, which are growing with each year. On further inspection, it was seen that the *SQL Injection* and *Cross-site Scripting* issues have visibly increased in recent years. Though, CWE-20 and CWE-476 have shown a considerable decrease till 2023, it has been considered under the increasing trend as the numbers have again started rising in 2024.

The final summary can be presented as follows:

> **Summary: With an overall decrease in automotive vulnerabilities across the last 7 years, there has been a noticeable shift in the CVE patterns from SoC/embedded to external applications. The CWE study justifies the same with an increasing trend of *SQL Injection* and *Cross-site Scripting* issues, which are more common for "software around" the vehicle rather than traditional embedded vehicle software.**

Having identified the frequently occurring CWEs in automotive software, it would be easier to trace their point of occurrence in the SDLC. Following the CWE descriptions provided by MITRE [15], it is possible to get an idea of the phase of origin of the CWE. This information should be further inspected and studied in light of other standards like the Open Worldwide Application Security Project (OWASP) [15]. This way, the SE team could map the automotive vulnerabilities to the SDLC phase and have a more holistic view of the secure software development process.

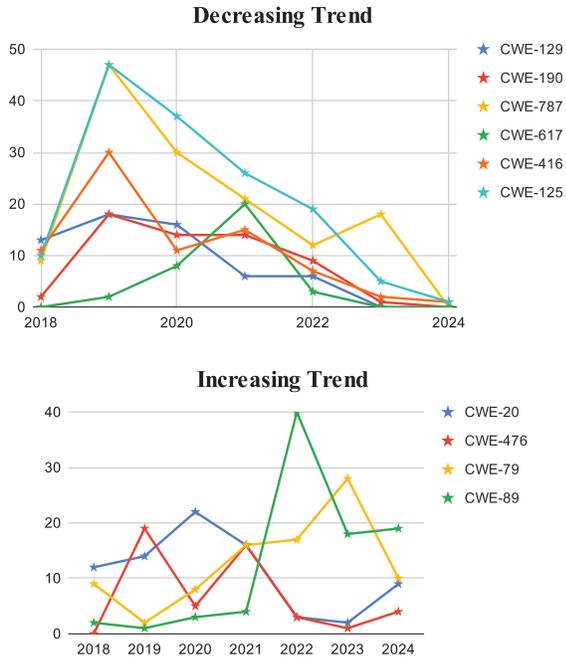

Fig. 8. Trend of CWE Counts (2018-2024)

## VI. THREATS TO VALIDITY

Some automotive vulnerabilities or attack vector mappings could go undetected using this paper's approach. We used the frameworks by Wohlin [17] and Staron [18] while considering the validity of our study.

In the *construct validity* category, the main risk was missing important CVEs when importing the CVE data from NVD. Vulnerability databases, like Vuldb, CISA, etc., can also be used. Another *construct validity* threat is using the CVSS score based on CVSS Version 3.0. However, we acknowledge that the results section could have been different if we had used version 2.0, 3.1, or 4.0. 3.0 is used here, as it is the most stable version that spans across the studied timeline

In the *conclusion validity* area, we see our main risk as our manual mapping for attack vectors. Although we considered using automated tools, e.g., LLMs, we chose to do it manually to be able to validate our results or discuss them if needed.

For the *external validity*, we consider the main risk to be the sample and population of our study. Software considered for automotive use may be a topic of further investigation.

## VII. CONCLUSION AND FUTURE WORK

The CV ecosystem comprising different automotive software is complex, especially from a security perspective. This study helps us understand the changing landscape of automotive vulnerabilities. This insight is important because an automotive SE team should know the nature and point of occurrence of these vulnerabilities. This would not only help them design secured software components but also know the correct place and time across the SDLC to prioritize and implement these designs.

We imported and filtered the automotive software vulnerabilities of the last seven years (2018 - September 2024) from NVD. 1663 vulnerabilities were reported, with the majority having **HIGH** CVSS score, and more variations in the CVSS value were found around 2023 and 2024. The attack vector mapping and CWE analysis both pointed to the fact that the automotive vulnerabilities have seen a major shift in trend from the year 2022, which was characterized by a decreasing number of SoC/Embedded system-related vulnerabilities and an increase in the vulnerabilities associated with external entities of the vehicle (e.g., web/mobile applications interacting with the vehicle, EV charging software, Car Rental/Dealership software, etc).

We have also discussed how this study can assist in mapping the vulnerabilities to the SDLC phase in which they might be introduced. We envision that our results will help software engineering teams improve the secure software development process for connected vehicles.

In the future, we plan to complete the SDLC phase mapping and refine the attack vector mapping process with an LLM-based model. Moreover, we plan to propose a prioritization strategy for these vulnerabilities that can benefit overall vulnerability management and mitigation.